# A network theoretic study of potential movement and spread of *Lantana camara* in Rajaji Tiger Reserve, India

Shashankaditya Upadhyay[1], Tamali Mondal[1], Prasad A. Pathak[2], Arijit Roy[3], Girish Agrawal[4], Sudeepto Bhattacharya[1*],

[1] Department of Mathematics, School of Natural Sciences, Shiv Nadar University, Post Office Shiv Nadar University, Gautam Buddha Nagar, Greater Noida 201 314, Uttar Pradesh, India

[2] Department of Physical and Natural Sciences, School of Liberal Studies, FLAME University, District Pune 412 115, Maharashtra, India

[3] Indian Institute of Remote Sensing, Indian Space Research Organization, 4, Kalidas Road, Dehradun 248 001, Uttarakhand, India

[4] Department of Civil Engineering, School of Engineering, Shiv Nadar University, Gautam Buddha Nagar, Greater Noida 201 314, Uttar Pradesh, India

[*]Corresponding author. Email: sudeepto.bhattacharya@snu.edu.in

## Abstract

Ecosystems are often under threat by invasive species which, through their invasion dynamics, create ecological networks to spread. We present preliminary results using a technique of GIS coupled with complex network analysis to model the movement and spread of *Lantana Camara* in Rajaji Tiger Reserve, India, where prey species are being affected because of habitat degradation due to Lantana invasion. Understanding spatio-temporal aspects of the spread mechanism are essential for better management in the region. The objective of the present study is to develop insight into some key characteristics of the regulatory mechanism for lantana spread inside RTR. Lantana mapping was carried out by field observations along multiple transects and plots and the data generated was used as input for MaxEnt modelling to identify land patches in the study area that are favourable for lantana growth. The patch information so obtained is integrated with a raster map generated by identifying different topographical features in the study area which are favourable for lantana growth. The integrated data is analysed with a complex network perspective, where relatively dense potential lantana distribution patches are considered as vertices, connected by relatively sparse lantana continuities, identified as edges. The network centrality analysis reveal key patches in the study area that play specialized roles in the spread of lantana in a large region. Hubs in the lantana network are primarily identified as dry seasonal river beds and their management is proposed as a vital strategy to contain lantana invasion. The lantana network is found to exhibit small-world architecture with a well formed community structure. We infer that the above properties of the lantana network have major contribution in regulating the rapid infestation and even spread of the plant through the entire region of study.

## 1. Introduction

India is home to a vast variety of flora and fauna, thriving or struggling to survive across its diverse landscapes comprising nine biogeographical zones [1]. A large number of exotic species originating elsewhere in the world have found sanctuary in India. *Lantana camara* is one such weed, a pan-tropic species that is thriving in many regions of the Indian subcontinent.

A non-native plant species, which upon colonization of some specific natural areas dominate the vegetation composition of these newly colonized areas, is known as invasive plant species. Invasive plant species may adversely affect the native biodiversity and may have a detrimental impact on the ecosystem processes of the colonized area [2]. Invasive plant species can modify nutrient cycling within the invaded environment resulting in altering the plant and animal populations [3][4][5]. However, there have been very few studies that have focused on invasive plants in tropical Asia and particularly in India, there is no systematic assessment of the extent of invasion within protected areas, the susceptibility of different habitats to invasion or the threats posed by invasive plants to native flora and fauna.

Rapid spread and colonization by invasive species is regarded as a major cause of global biodiversity depletion [6]. *Lantana camara*, being regarded as one of the most successful invaders in India is often seen as a contemptible threat to native biodiversity [7][8][9][10][11]. Lantana was introduced to India by the British in 1809 as an ornamental plant [12][13][14]. At present, it is found in almost all the regions of the subcontinent barring Thar Desert and its surrounding areas [13][15][16][17][18][19]. The regions in India where lantana is present are witnessing a sharp increase in the density of its infection while at the same time lantana is rapidly colonizing previously unblemished geographic domains in India [20][21][22][23][24].

Lantana is often found to be toxic to other plants and eliminate native vegetative species through smothering and negative allelopathic effects [25][26][27]. The dense thickets of lantana may aid in elevating the intensity of wild fires, particularly along the forest edges [28][29]. As a threat to agriculture lantana invades pastures and grazing lands, riparian areas, cultivated land and orchards thus increasing the costs of production by impeding access for vehicles, machinery and stock and reducing fertility of land. It is toxic to livestock [30][31].

The primary objective of our study is to model the spread of lantana as a complex system. Researchers have tried to present mechanisms of spread of lantana in the past. However, we wish to use complex networks as a method to study the spread of lantana across a study area. Networks are modern day formal graph-theoretic tools to model empirical data from real world systems that can be conceived as an assembly of interactive components such that the quality of interaction is of interest [32]. The methods being formal can be extended to model other systems of invasion in an ecosystem which have similar dynamics of growth and spread.

As a formal and contemporary paradigm, use of mathematical theory of complex networks may enhance our understanding of regulatory mechanisms of lantana spread and growth across a given region. The generated knowledge of spread and regulatory mechanism of lantana may help develop proper management strategies to tackle lantana spread in case it poses a serious threat to a native ecosystem. In our understanding, applying the theory of complex networks to address the mechanisms of spread of in this region of the world promises to present us with an opportunity to perceive insights into the spread and regulatory mechanisms of lantana. [33].

## 2. Study area

Rajaji National Park, comprising of a total geographic area of 1075 sq. km, has been declared as Rajaji Tiger Reserve (RTR) by National Tiger Conservation Authority (NTCA), Government of India in 2015. The park encompasses three former wildlife sanctuaries namely, Rajaji, Motichur and Chilla as well as part of the Shivalik and Dehradun East forest divisions in the state of Uttrakhand.

Chilla range within Rajaji Tiger Reserve formed the intensive study area for the present work. Chilla range is spread between the districts of Dehradun and Haridwar. Chilla lies between N29°50' to N30°50' latitude and E78°10' to E78°25' longitude, comprises of 171.678 sq. km area of the reserve connects the Western portion of RTR, located west of the Ganges, with the Shivalik landscape to the east. The geographic location of the study area is highlighted in Fig. 1.

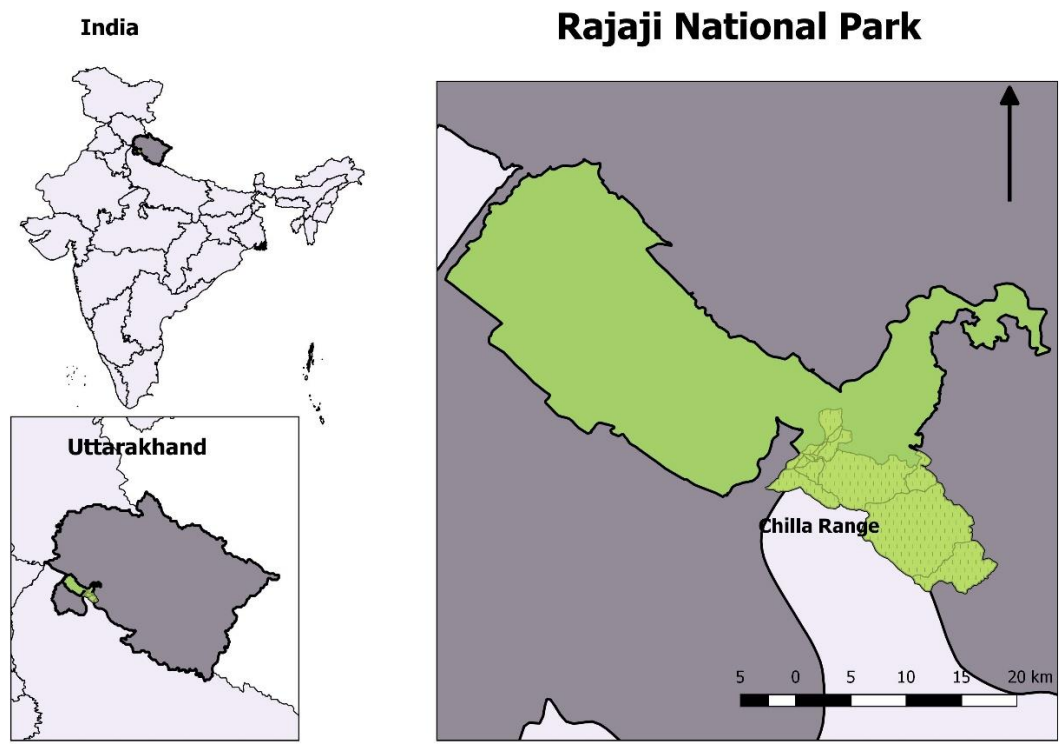

Fig. 1: Geographic location of the study area. Highlighted area in brown on the map of India is the state of Uttarakhand. Within Uttarakhand, the highlighted area in green is the Rajaji National Park and the dotted region within is the Chilla range shown here.

The climate of RTR is subtropical which is slightly moderated due to its vicinity to the outer Himalayas. The temperature in the region shows an average high of around 42° C in the months of May-June and can fall below 2° C during the months of December-January, which are marked by a frequent presence of fog and frost. The annual rainfall in the region varies from 1300mm in the outer hills to 1900mm in the upper hills [34], with generally mid-July being the period of onset of monsoon resulting in a moisture laden atmosphere usually lasting till the end of October.

Lantana is widely present in many parts of RTR as the conditions in these areas are favourable for the growth of lantana. The plant is found at altitudes from sea level to 2,000 m and can thrive very well under rainfall ranging from 750 to 5000 mm per annum. Lantana growth is reduced when the surrounding temperature falls below 5° C. At places where natural forests have been disturbed through logging that results in creation of canopy-free regions or gaps, lantana encroaches in the gaps. Preferable conditions for lantana growth are availability of light and moisture in soil throughout the year. It can tolerate partial shading and prolonged dry periods but dies as a result of water logging, salinity or complete shading through canopy cover. Thus lantana does not invade intact rainforests but is found on their margins. The various biological components of lantana are shown in Fig. 2. A general summery of the environmental preference of the plant is given in Table 1.

Table 1. A brief description of habitat, soil requirement and dispersal of *Lantana camara*.

| Habitat | Soil aspect | Flowering | Pollination | Seed dispersal and germination |
|---|---|---|---|---|
| Lantana grows in a variety of coastal and subcoastal areas, thriving in high rainfall areas of tropical, subtropical and warm temperate climates. | It grows well in rich organic soils, well-drained clay soils, and volcanic soils derived from basalts but will also tolerate poor soils and almost pure sands, as long as moisture is available. It will not grow in the tropics if soils are shallow and have a very limited water-holding capacity [35]. | Flowering occurs between August and March, or all year round if adequate moisture and light is available. | Insects such as butterflies, moths, bees and thrips pollinate the flower clusters. Self-pollination is not common [36][37][38]. | Fruit-eating birds are the main agents of dispersal. Some mammals also disperse lantana seed. Seeds need warm temperatures and sufficient moisture to germinate. Germination is reduced by low light conditions [39]. |

Lantana as a species is very resilient and plants tend to die only under very stressful conditions. Extended droughts and complete shading through canopy closure are few known examples of such extremely stressful conditions that result in declination of lantana spread [36].

## 3. Material and methods

In order to obtain an insight into the extent of presence of lantana in RTR, field surveys were conducted in the study area exclusively during the period of January to June in the year 2016. The methodology of data collection was designed keeping in mind the block-subdivision of the study area. The Chilla range is subdivided into fifteen blocks, each of which was surveyed exactly once in order to avoid any sampling bias. A total of 15 transects of approximately 2 km in length, one in each of the blocks were identified such that the start and end positions of each transact was recorded using Garmin GPS.

On each of these transects, three quadrant plots of size 12 m by 12 m were laid out at the beginning, midway and the end of the transect and their GPS positions were recorded. Each quadrant plot was subdivided into four subplots of equal area and lantana presence-only data was collected based on presence of lantana in each subplot. Presence of lantana in only one of the subplots was identified as 0-25% presence. Likewise, presence of lantana was classified as 25-50%, 50-75% and 75-100% depending on presence of lantana in exactly two, three and all of the four subplots respectively.

The presence-only data thus collected formed the basis of generating a species distribution model for *Lantana camara* in the study area using the maximum entropy (MaxEnt) algorithm. MaxEnt determines the best potential distribution by selecting the most uniform distribution subject to the constraint that each environmental variable in the modelled distribution matches its empirical average over the known distributional data [40]. Additionally, since canopy free regions are conducive to lantana growth, canopy-less regions in the study area (including river belts, seasonal streams, grasslands etc.) were identified and subsequently digitised Google Earth, Cartosat 1 Digital Elevation Model version 3 data with spatial resolution 30 m dated 4 July 2016 from Earth Explorer (http://earthexplorer.usgs.gov) and data for boundaries, range and compartments of protected areas made available by Wildlife Institute of India, Dehradun. A resistance raster surface map for the study area was thus obtained with information on elevation, slope and aspect using GIS. The information on elevation, aspect and slope was used to configure environmental variables in MaxEnt. A Google Earth image with digitized river streams in the study area is shown in Fig. 3. for example.

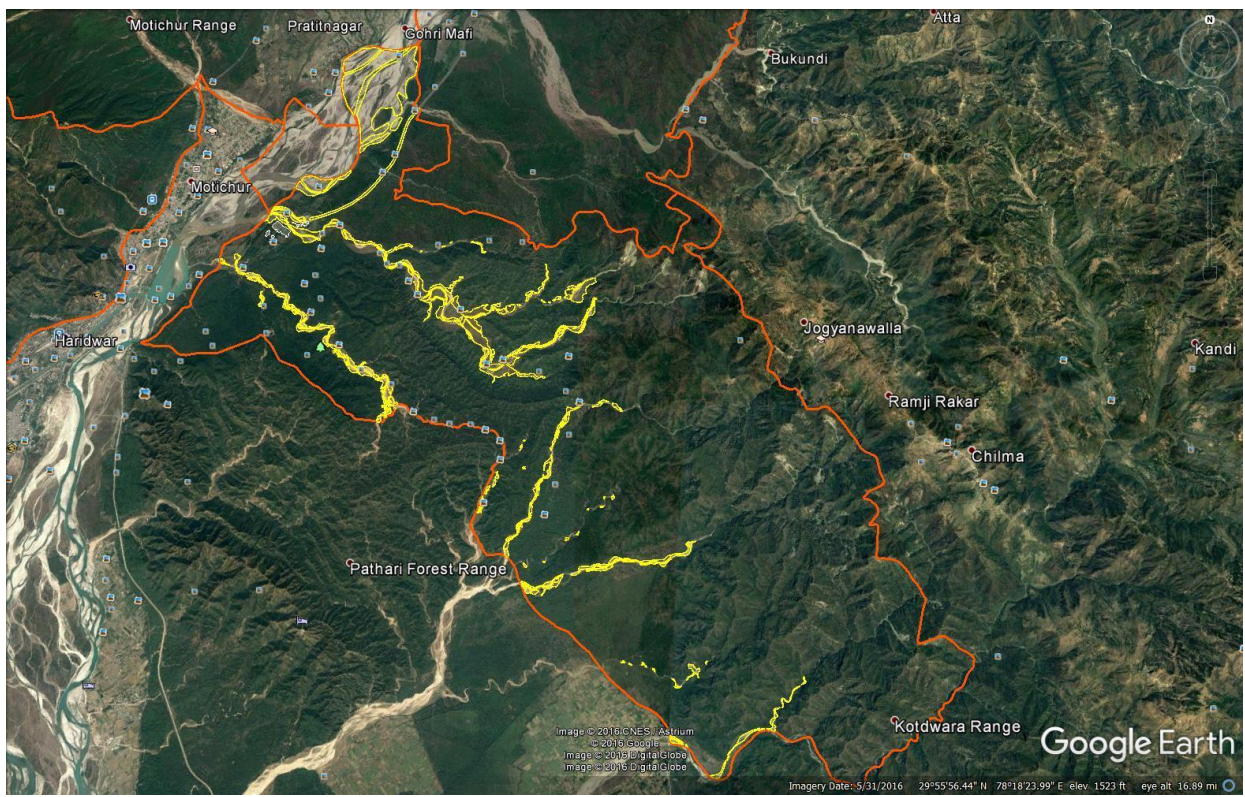

Fig. 2. Google Earth image showing digitized rivers and seasonal streams in the study area.

A functional network-theoretic model for potential spatial distribution and movement of Lantana among habitat patches in the study area was obtained using MaxEnt modelling and raster surface data. The vertices in this network model are patches of forest gaps (favourable for lantana growth) and they are located on the raster map. Vertices are connected by edges that take the least-cost path along the resistance raster surface. For our selected connectivity metrics, the maximum distance of travel between any two vertices is modelled at 700 m. Thus a network is obtained with a total of 808 vertices of patches where lantana is potentially distributed in the study area. The vertex with highest area is a potential distribution patch spread in 372.69 ha whereas the one with smallest area is 900 sq. m. A pictorial representation of the derived lantana network is given as Fig. 4.

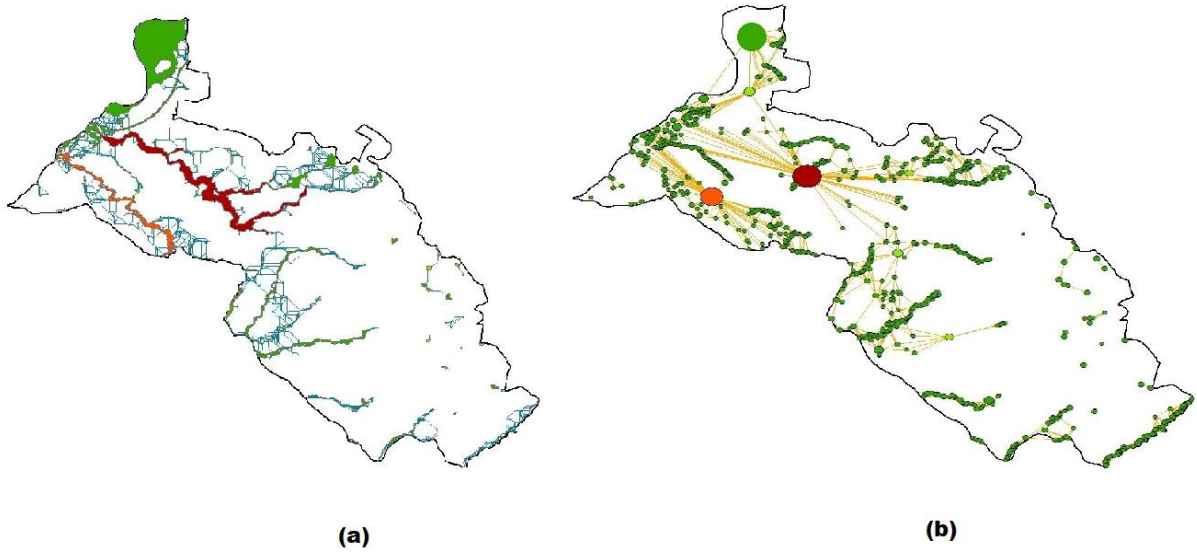

(a) (b)

Fig. 3. (a) Contiguous patches of potential lantana distribution connected by blue lines representing the least-cost paths between two patches. (b) Lantana network showing patches as nodes and least-cost paths as edges. The patches with largest, second-largest and third largest area are coloured as green, red and orange respectively.

3.1 Network centrality

Centrality indices are explicitly used to quantify the notion that by the virtue of their structure and orientation some vertices in the network play a more central and important role in the various process occurring across the network. A real valued function is called a centrality index if and only if the following conditions are satisfied

$\forall x, y, z \in X, f(x) \leq f(y) \text{ or } f(y) \leq f(x)$

$f(x) \leq f(y) \text{ and } f(y) \leq f(z) \text{ then } f(x) \leq f(z)$

$f(x) \leq f(y) \text{ and } f(y) \leq f(x) \text{ then } f(y) = f(x)$ (1)

that is $f$ induces a total order on $X$. By this order we can say that $x \in X$ is at least as central as $y \in X$ with respect to a given centrality $c$ if $c(x) \geq c(y)$ [41].

Over the years, researchers have proposed various centrality indices useful in a variety of applications; some more than others depending on the nature of the problem under investigation. Some of these indices are discussed here.

The degree of a vertex u in a graph G is the number of edges incident on that vertex. The degree centrality of a vertex is given by

$$C_D(u) = d(u) \qquad (2)$$

where d(u) is the degree of the given vertex [42]. It can be computed as marginal of adjacency matrix A

$$C_D(i) = \Sigma_j A_{ij} x_j = (Ax)_i \qquad (3)$$

where $x$ is a vector with each entry as 1.

As per degree centrality a vertex is more central the more it is directly connected to other vertices. Thus degree centrality is understood as a measure of immediate influence in a network.

Eigenvector centrality conveys the idea that a vertex is more central as compared to other vertices in a network central if either it has many adjacent vertices or it is adjacent to vertices that have many adjacent vertices (and hence are more central in the network) [43]. Eigenvector centrality is given by the equation

$$Ax = \lambda_1 x \qquad (4)$$

where $\lambda_1$ is the largest eigenvalue of adjacency matrix A of the network.

Betweenness centrality characterizes the importance of a vertex when information is passed through a given pair of vertices [44]. Betweenness centrality measures the number of times the information passes through a vertex k by a shortest path between vertices i and j. Let $\delta_{st}(v)$ denotes the fraction of shortest paths between vertices s and t that contains v such that $\delta_{st}(v) = \frac{\sigma_{st}(v)}{\sigma_{st}}$, where $\sigma_{st}$ is the number of all shortest path between s and t. The betweenness centrality of a vertex is given by

$$C_B(v) = \Sigma_{s \neq t \neq v} \delta_{st}(v) \qquad (5)$$

Closeness Centrality considers a vertex more central if the total sum of distance from the given vertex to all other vertices is minimum. The closeness centrality is given by

$$C_C(u) = \frac{1}{\Sigma_{v \in V} d(u,v)}. \qquad (6)$$

While closeness centrality can be easily computed for a connected graph, it is not defined for a graph with multiple connected components since the distance between two vertices belonging to different components in a graph is considered as infinite. To address this issue a different form of distance based centrality called harmonic centrality has been proposed [45].

Harmonic centrality calculates the sum of reciprocal of distances instead of the reciprocal of sum of distances as in the case of closeness centrality. Formally, harmonic centrality is defined as

$$C_H(u) = \sum_{u \neq v} \frac{1}{d(u,v)}. \tag{7}$$

The measure is useful in the applications where a vertex is ranked higher if the vertex is close to most other vertices in the network (in terms of geodesic distance).

3.2 Identification of small-world networks

Clustering coefficient of a vertex in a network is the ratio of number of edges present among the adjacent vertices of the vertex and total number of edges possible among the adjacent vertices. Let $G = (V,E)$ be a graph. The clustering coefficient of a vertex $u$ with set of adjacent vertices $N_u$ is given by

$$C(u) = \frac{2 \mid e_{ij} : i,j \in N_u, e_{ij} \in E \mid}{k\ (k-1)} \tag{8}$$

Where k is the number of vertices in $N_u$.

A similar idea is that of transitivity, which is the ratio of three times the number of triangles present in a graph to the number of connected triplets (L-shaped formations) of vertices in the graph [46]. Transitivity is defined as

$$T = \frac{3 \times number\ of\ triangles}{number\ of\ paths\ of\ length\ 2} \tag{9}$$

A network is said to possess small-world property if any vertex in the network can be reached by any other vertex in the network by traversing a path consisting of only a small number of vertices. It is observed that for a network $G$ of order $n$ and size $m$, the average shortest distance $ASD$ is similar to an Erdos - Rényi random graph of same size and order. But the transitivity $T_G$ and clustering coefficient $C_G$ of the network is much higher than that for the Erdos - Rényi random graph $T_{ER}$ and $C_{ER}$ [46]. This property is used to define a measure of small-world property. We call this as *small-world-ness* [47], which is defined as

$$SW = \frac{T_G \times L_{ER}}{L_G \times T_{ER}} \tag{10}$$

where $L_G$ and $L_{ER}$ are average shortest path length for the given network $G$ and Erdos - Rényi random graph of same size and order. We call a network to be small-world if the value of small-world-ness is greater than one.

3.3 Assortative mixing

A network is said to show assortative mixing if vertices of higher degree are connected to vertices of higher degree and the vertices of lower degree tend to connect to vertices of lower degree. A network is called disassortative if the high degree vertices are connected to vertices of low degree. A measure for assortative mixing has been proposed [48]. In a graph with $N$ vertices and $M$ edges suppose we choose an edge $e$ and arrive at a vertex $v$ along the chosen edge $e$. Then the distribution of the remaining degree (the number of edges leaving the vertex other than the one along which one arrived) is given by

$$q_k = \frac{(k+1)p_{k+1}}{\Sigma_j\, j\, p_j} \tag{11}$$

where $p_k$ is the probability that a vertex chosen at random in a network has degree equal to $k$.

Let $e_{jk}$ be the joint probability distribution of the remaining degrees of the two vertices at either end of a randomly chosen edge. Then the assortative mixing ($r$) is given by

$$r = \frac{1}{\sigma_q^2} \Sigma_{jk}(e_{jk} - q_j q_k) \tag{12}$$

where $\sigma_q^2 = \Sigma_k k^2 q_k - [\Sigma_k k\, q_k]^2$ is the variation of $q_k$.

For the purpose of computations, assortative mixing is calculated as

$$r = \frac{M^{-1} \Sigma_i\, j_i\, k_i - [\,M^{-1} \Sigma_i \frac{1}{2}(j_i + k_i)]^2}{M^{-1} \Sigma_i \frac{1}{2}(j_i^2 + k_i^2) - [M^{-1} \Sigma_i \frac{1}{2}(j_i + k_i)]^2} \tag{13}$$

where $j_i$ and $k_i$ are the degrees of the vertices at the end of the $i$th edge where $i = 1 \ldots M$.

3.4 Communities in networks

Communities in a network are set of vertices that are present such that they have more edges linking the vertices within the set as compared to vertices outside of the set. Which means that the set forms a local cluster and has a locally cliquish topology. An arbitrary partition of the vertex set of a graph is said to form a good community structure if the value of modularity associated with it is close to one. Modularity as a measure was first proposed by Newman and Girvan in their seminal work on community detection [49]. A partition of a network into $k$ communities can be represented by a $k \times k$ symmetric matrix $E$ whose element $E_{ij}$ is the fraction of all the edges connecting vertices in community $i$ with vertices in community $j$. Given such a matrix $E$, modularity is defined as

$$Q = \Sigma_i(E_{ii} - a_i^2) = Tr(E) - ||\,E^2\,|| \tag{14}$$

where $a_i = \Sigma_j E_{ij}$ represent the fraction of edges with one end in community $i$, $Tr(E)$ is trace of matrix $E$ and $||\,A\,||$ is the sum of elements of matrix $A$. Clearly, the maximum value possible for $Q$ for any partition is one. Newman further proposed a fast greedy algorithm for community detection which starts with considering each vertex as a community and iteratively merges two communities such that a maximum level of increase in the value of modularity is achieved on merging the two communities [50].

# 4. Results

The network of potential lantana distribution patches or lantana network consist of a total of 808 vertices or lantana patches. Of these 808 potential lantana distribution patches, only five patches are isolated or not contiguous i.e. the degree of these vertices in the network is zero. Thus with a mean degree of 4.74, almost all patches present in lantana network are in vicinity of some other lantana patch and thus contiguous.

The various centrality indices are computed for the lantana network and the ten best ranked potential lantana distribution patches as ranked by the centrality indices and the values of these indices are given here as Table 2.

| Degree Centrality ($C_D$) | | Eigenvector Centrality ($C_E$) | | Betweenness Centrality ($C_B$) | | Harmonic Centrality ($C_H$) | |
|---|---|---|---|---|---|---|---|
| Vertex No. | Degree | Vertex No. | Value | Vertex No. | Value | Vertex No. | Value |
| 65 | 72 | 65 | 0.6345 | 65 | 0.4489 | 65 | 202.26 |
| 124 | 61 | 11 | 0.1346 | 458 | 0.2279 | 124 | 167.50 |
| 11 | 27 | 240 | 0.1237 | 124 | 0.2261 | 11 | 167.46 |
| 47 | 17 | 267 | 0.1060 | 468 | 0.2225 | 55 | 157.44 |
| 240 | 16 | 290 | 0.1045 | 70 | 0.2111 | 240 | 154.13 |
| 1 | 15 | 299 | 0.1026 | 78 | 0.1905 | 67 | 152.29 |
| 533 | 14 | 55 | 0.1004 | 55 | 0.1809 | 458 | 149.96 |
| 148 | 13 | 317 | 0.0983 | 533 | 0.1599 | 64 | 149.93 |
| 660 | 12 | 381 | 0.0975 | 465 | 0.1095 | 70 | 148.14 |
| 27 | 11 | 67 | 0.0970 | 464 | 0.1052 | 47 | 148.10 |

Table 2. The ten best ranked potential lantana distribution patches as ranked by different centrality indices

The aforementioned best ranked patches of potential lantana distribution, ranked as per the different centrality indices used, are identified on the map and the resulting network visualization of these best ranked patches is presented as Fig. 4 and Fig. 5.

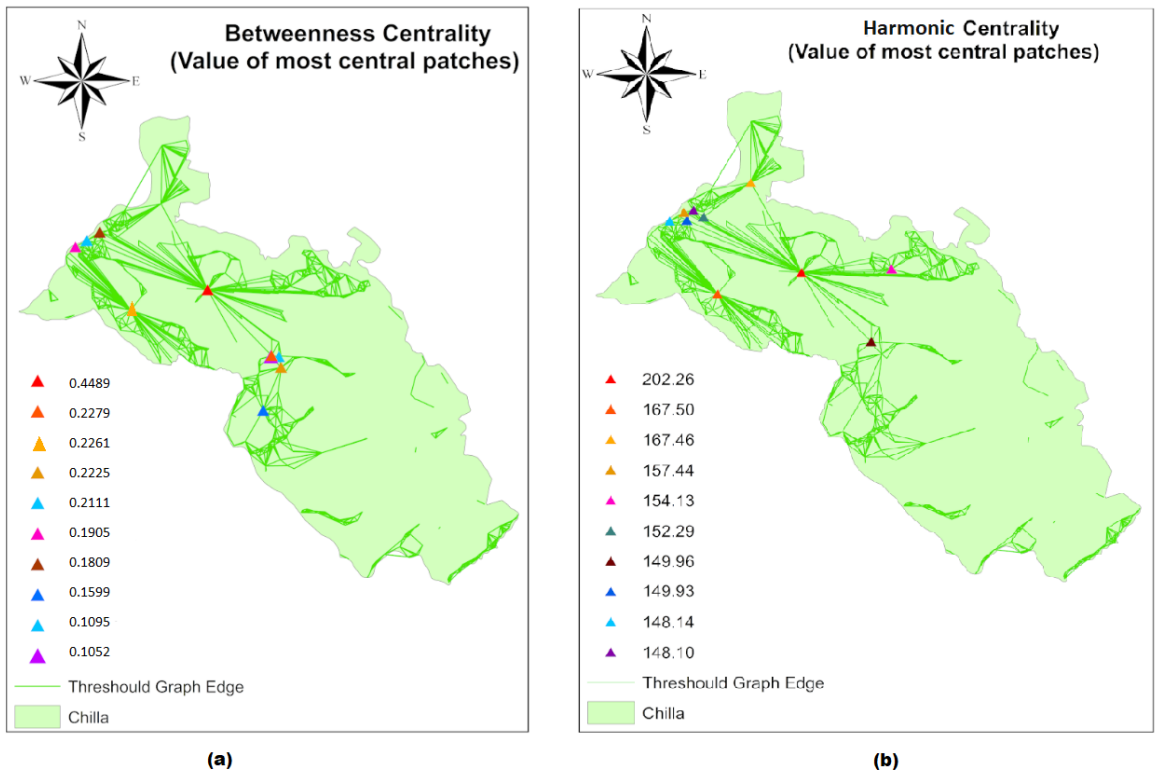

Fig. 4. The 10 best ranked potential lantana distribution patches as ranked by (a) betweenness centrality and (b) harmonic centrality.

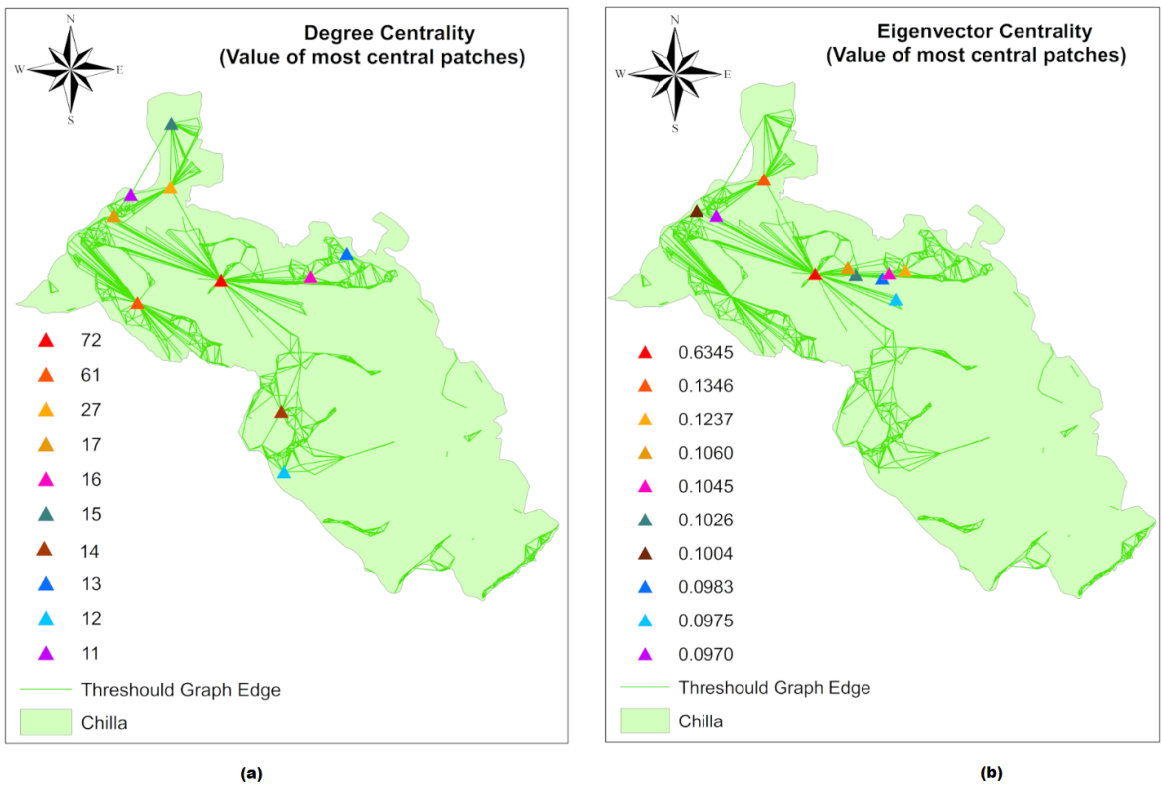

Fig. 5. The 10 best ranked potential lantana distribution patches as ranked by (a) degree centrality and (b) eigenvector centrality.

For the lantana network the value of clustering coefficient and transitivity is found to be equal to 0.5280 and 0.2672 respectively. The average shortest distance of the network is 5.1992. The average shortest distance and transitivity for an Erdos – Renyi random network of same size and order is found to be equal to 4.3639 and 0.0053 respectively. Thus small-world-ness of the network is equal to 42.6931. Thus the lanata network is a small-world network.

The value of assortativity in lantana network is calculated to be -0.0704, which is very close to zero. The lantana network thus does not show assortative mixing. Moreover, the community structure of lantana network is found to be composed of 27 mutually overlapping large and small communities. The largest community consists of 156 vertices or patches of forest gaps favourable for lantana growth while the smallest ones are five isolated vertices or forest gaps, which are not in vicinity of other forest gaps where lantana might be present. The overlapping community structure in lantana network is well formed as indicated by the value of modularity, which is calculated to be 0.8285. A network visualization of the community structure showing a total of nine communities, with at least thirty potential lantana distribution patches in each community as its members, is represented in Fig. 6. The member vertices of each community are represented by a different colour.

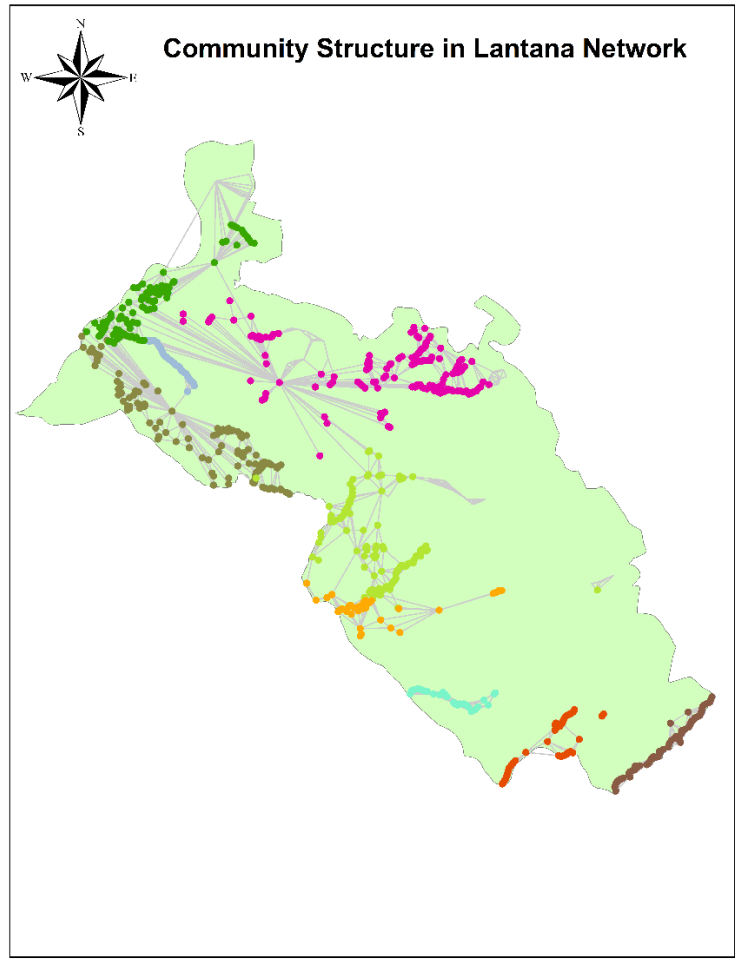

Fig. 6. Community Structure of lantana network showing nine communities, each represented by a different colour, all of which contain at least thirty potential lantana distribution patches as its members.

## 5. Discussion

The patches of potential lantana distribution that are top-ranked as per the different centrality indices may be considered important centres for the movement and spread of lantana in the study area. However, the ecological role played by a patch ranked high by a centrality index may not be the same as some other central patch which is ranked higher as per some different centrality index. Thus the patches of potential distribution of lantana that are singularly ranked high as per some centrality index may play a specific role in the growth and spread of lantana in the studied region.

Hubs in a network are vertices with degree far higher than the average value. The central lantana patches that are ranked among the top by the degree centrality can be considered as hubs in the network. These hub lantana patches which are close to a large number of patches that are favourable for lantana growth are very important centres in the dynamics of lantana growth in the studied region. If lantana is present in one of the hub patch then it can potentially spread to a large number of adjacent patches.

As indicated by the lantana network being non-assortative, hubs in the network may or may not always be adjacent. The vertices in a network which are ranked higher by the eigenvector centrality are either hubs or vertices that are adjacent to some non-adjacent hub vertices. Thus some potential distribution patches that are ranked higher by the eigenvector centrality but are not hub patches gain importance in the network. These patches may act as relay centres for the spread of lantana between hub patches.

It is observed that the patches that are best ranked according to the betweenness centrality in given network are those patches favourable for lantana growth that form the few vertices that are adjacent to members of different communities. Thus their role in the spread of lantana in the region becomes very crucial as lantana originating in these patches can occupy a large number of vertices or favourable patches in adjoining communities. Likewise, the patches that are ranked higher as per the harmonic centrality are the ones that can boost the spread of lantana to a large number of patches to which they are close in terms of graph distance i.e. lantana originating in these patches can reach a large number of patches following a few edges or least-cost paths.

On plotting the highest degree vertices or potential lantana distribution patches on the map, we observe that these vertices of higher degree or hubs are almost always the dry seasonal river beds in the study area. Thus by the virtue of their topography, these dry river beds falls in the vicinity of a large number of patches that are favourable for lantana growth. At the same time, since the soil content of these patches is highly conducive for vegetation growth and are mostly without a thick forest cover, we deduce that these dry river beds form the most critical base for the spread of lantana in a region. An example of such dry river beds being favourable for lantana growth is reflected by unanimously high rank being given to patch number 65 and 124 by different centrality indices. Both these patches are part of seasonal dry river beds and are individually the second-largest and third-largest potential lantana distribution patches respectively by area.

The lantana network being a small-world network with a well formed community structure implies that even being sparsely connected, there is high clustering at the local scale together with short average path lengths. For the process of spread of lantana in the region this would mean that through the favourable patches in which it is evenly present, lantana will propagate to neighbouring patches and since the neighbouring patches are themselves neighbours (presence of triangles in the network) the movement of lantana will take place evenly and fast on a local scale.

As a consequence of the small-world property and a well-established community structure in lantana network, an interesting speculation can be made for the dynamics of movement and spread of lantana when it would have first invaded a connected component of network of patches favourable for lantana growth studied here. The inferred dynamics shall be applicable even today when lantana continue to grow in the studied region. Due to clustering, lantana when first conquering a favourable patch will showcase an even spread locally in the connected component as discussed earlier. Also, due to low average shortest path length in the network the spread of lantana becomes possible to relatively distant patches in the component as all the patches are only a few edges away from each other. Thus due to mesoscale transitions the spread of lantana happens rapidly in a component on an ecological time scale. Moreover, if a hub patch is present in the component then there is a sudden burst in the spread of lantana when the species conquers a hub patch and begins to propagate to all its adjacent patches. Thus lantana when introduced to a connected component is likely to establish its presence across the entire component because of presence of communities and the small-world nature of the network. Furthermore, the fact that zoochory and ornithochory are the primary mechanisms for seed dispersal of lantana only adds to its menace as it is likely that lantana is introduced in a favourable patch despite the patch being surrounded by thick canopy-covered regions.

Informed by our study, we opine that a viable strategy to tackle lantana invasion in the study area before it becomes a reasonable threat is to first manage the most central patches in the region as a reduced activity of lantana movement through these most central patches can significantly lower the rate of movement and spread of lantana to other patches. However, since the network of forest patch showcase high clustering, it is very likely that lantana when removed from a patch can re-engulf it from its presence in the neighbouring patches. Thus intense reforestation is a highly recommendable strategy to prevent the movement of lantana in the entire region which, however slow, may help contain the lantana spread in a larger region.

## 6. Conclusion

Network theoretic modelling of spread of *lantana camara* in the region of study helps us identify some key areas in the region that are critical for the movement of spread of lantana. These critical areas, identified using various centrality measures, play specialized roles in the dynamics of spread of lantana in the region under investigation. The identification of these central areas for the growth of lantana may provide the managers and ecologists with initial starting points to schematically curtail through proper land management and policies the growth of lantana in the region, if ever it poses serious threats to the native environment and ecology.

Hub patches in lantana network are identified as areas in the region of study which primarily are parts of dry seasonal river beds. In order to diminish the rate of growth of lantana in the region, it is necessary that these hub patches are properly managed by the wildlife authority.

Presence of locally dense connections in the network is deduced as indicated by high clustering and presence of well-formed community structure in the network. Together with an average shortest path length comparable to a random network, the observed high clustering classifies the lantana network as a small-world network. We thus conclude that lantana propagate rapidly and evenly on a local level in the network of favourable patches due to small-world property in the network. At the same time we also infer that due to these global properties of the network it is difficult to eliminate lantana infection from a patch as it can easily get re-infected if lantana is present in a neighbouring patch. Thus managing lantana invasion in the studied region may not be an easy process.

Finally, it must be stated that if a similar study is conducted in some other region of interest, the global properties of network of patches favourable for lantana growth in that region shall depend on the topographical, geographical and ecological properties of that region which may not be same as the global properties of network studied here. However, since our methods are formal and definitive we conclude that wherever these global network properties are same as network studied here, the spread and growth of lantana in that region shall follow a similar dynamics as proposed in this study.

## 6. Acknowledgements

We would like to thank the Field Director, Rajaji Tiger Reserve, for allowing us to carry out the surveys in Chilla range, RTR.

## 7. References

[1] Mani, M. S., 1974. Ecology and Biogeography in India, 10.1007/978-94-010-2331-3.

[2] Colautti, R.I., MacIsaac, H.J., 2004. A neutral terminology to define 'invasive species'. Diversity and Distributions, 10: 135–141.

[3] Didham, R.K., Tylianakis, J.M., Hutchison, M.A., Ewers, R.M. and Gemmell, N.J., 2005. Are invasive species the drivers of ecological change? Trends in Ecology & Evolution, 20(9): 470-474.

[4] Didham, R.K., Tylianakis, J.M., Gemmell, N.J., Rand, T.A. and Ewers, R.M., 2007. Interactive effects of habitat modification and species invasion on native species decline. Trends in Ecology & Evolution, 22(9): 489-496.

[5] Strayer, D.L., Eviner, V.T., Jeschke, J.M., and Pace, M.L., 2006. Understanding the long-term effects of species invasions. Trends in Ecology and Evolution, 21: 645–651.

[6] Wilcove, D. S., and Chen, L.Y., 1998. Management Costs for Endangered Species. Conservation Biology, 12: 1405–1407.

[7] Murali, K.S., Setty, R.S., 2001. Effects of weeds Lantana camara and Chromelina odorata growth on the species diversity, regeneration and stem density of tree and shrub layer in BRT sanctuary, Current Science, 80: 675-78.

[8] Sharma, G.P., and Raghubanshi, A.S., 2006. Tree population structure, regeneration and expected future composition at different levels of Lantana camara L. invasion in the Vindhyan tropical dry deciduous forest of India. Lyonia, 11(1): 27-39.

[9] Prasad, A.M., Iverson, L.R.., Liaw, A., 2006. Newer classification and regression techniques: bagging and random forests for ecological prediction. Ecosystems, 9: 181–199.

[10] Sahu, P.K., Singh, J.S., 2008. Structural attributes of lantana-invaded forest plots in Achanakmar–Amarkantak Biosphere Reserve, Central India. Current Science, 94(4): 494–500.

[11] Babu, S., Love, A., Babu, C.R., 2009. Ecological restoration of lantana-invaded landscapes in Corbett Tiger Reserve, India, Ecological Restoration, 27: 467–477.

[12] Brandis, D., 1882. Suggestions Regarding Forest Administration in the North-Western Provinces and Oudh: Including a Joint Report on the Forests of the School Circle. Home Department Press.

[13] Aravind, N.A., Rao, D., Vanaraj, G., Ganeshaiah, K.N., Shaanker, R.U., Poulsen, J.G., 2006. Impact of *Lantana camara* on plant communities at Malé Mahadeshwara reserve forest, South India. In: L.C. Rai and J.P. Gaur (eds.), Invasive alien species and biodiversity in India. Banarus Hindu University. Banarus. India, 68–154.

[14] Nanjappa, H.V., Saravanane, P., Ramachandrappa, B.K., 2005. Biology and management of Lantana camara L.–a review. Agricultural Reviews, 26(4): 272-280.

[15] Kannan, R., Aravind, N.A., Joseph, G., Ganeshaiah, K.N., Shaanker, R.U., 2008. Lantana Craft: A Weed for a Need. Biotech News, 3(2): 9-11.

[16] Kimothi, M.M., Dasari A., 2010. Methodology to map the spread of an invasive plant (*Lantana camara* L.) in forest ecosystems using Indian remote sensing satellite data. International Journal of Remote Sensing, 31(11-12): 3273–3289.

[17] Surampalli, D., Abundance of Lantana camara in Open Canopy, Partial Canopy, Closed Canopy Areas in Forest Trails, Karnataka, India.

[18] Patel, J., Kumar, G.S., Deviprasad, S.P., Deepika, S., Qureshi, M.S., 2011. Phytochemical and anthelmintic evaluation of Lantana camara (L.) Var. Aculeate leaves against pheretima posthuma. Journal of Global Trends in Pharmaceutical Sciences, 2(1): 11-20.

[19] Dobhal, P.K., 2010. Ecological Audit of Invasive Weed *Lantana camara* L. along an Altitudinal Gradient in Pauri Garhwal (Uttranchal). Ph.D. dissertation, Punjab University, Chandigarh, India.

[20] Roy, P.S., Dutt, C.B.S., Joshi, P.K., 2002. Tropical forest assessment and management. Tropical Ecology, 43: 21–38.

[21] Day, M.D., Wiley, C.J., Playford, J., Zalucki, M.P., 2003. Lantana: Current Management, Status and Future Prospects. Australian Centre for International Agricultural Research, 5: 1- 20.

[22] Sharma, G.P., Raghubanshi, A.S., Singh, J.S., 2005. Lantana invasion: An overview. Weed Biology Management, 5: 157–167.

[23] Kohli, R.K., Batish, D.R., Singh, H.P., Dogra, K.S., 2006. Status, invasiveness and environmental threats of three tropical American invasive weeds (Parthenium hysterophorus L., Ageratum conyzoides L., *Lantana camara* L.) in India. Biological Invasion, 8: 1501-1510.

[24] Dogra K.S, Kohli R.K., Sood S.K., 2009. An assessment and impact of three invasive species in the Shivalik hills of Himachal Pradesh, India. International Journal of Biodiversity and Conservation, 1(1): 4–10.

[25] J. R. Qasem J. R. and Foy C. L., 2001. Weed Allelopathy, Its Ecological Impacts and Future Prospects, Journal of Crop Production, 4(2): 43-119.

[26] Ahmed, R., Uddin, M.B., Khan, M.A.S.A., Sharif, A. M., Mohammed, K. H. 2007. Allelopathic effects of *Lantana camara* on germination and growth behavior of some agricultural crops in Bangladesh. Journal of Forest Research, 18: 301-304.

[27] Sharma, O.P., Makkar, H.P.S., Dawra R.K., 1988. A review of the noxious plant Lantana camara, Toxicon, 26(11): 975-987.

[28] Berry, Z.C., Wevill, K., Curran, T.J., 2011. The invasive weed Lantana camara increases fire risk in dry rainforest by altering fuel beds. Weed Research, 51: 525-533.

[29] Ankila H., Bharath, S., 2005. The Fire-Lantana Cycle Hypothesis in Indian Forests. Conservation and Society, 3: 26-42.

[30] Sharma, O.P., Makkar, H.P.S., Dawra R.K., Negi S.S., 1981. A Review of the Toxicity of Lantana camara (Linn) in Animals. Clinical Toxicology, 18(9): 1077-1094.

[31] Fourie, N., Van der Lugt, J.J., Newsholme, S.J., Nel, P. W., 1987. Acute Lantana camara toxicity in cattle. Journal of the South African Veterinary Association, 58(4): 173-178.

[32] Newman, M.E.J. Networks: An Introduction, Oxford University Press, 2010.

[33] Upadhyay, S., Pathak, P.A., Agrawal, G., Bhattacharya, S., 2017. A network-theoretic modelling of spatial distribution of Lantana camara in Rajaji Tiger Reserve, India. ISEM Global Conference, Jeju, South Korea.

[34] Williams, A.C., Asir J.T., Johnsingh, A.J.T., Krausman, P.R., 2001. Elephant-Human Conflicts in Rajaji National Park, NorthWestern India. Wildlife Society Bulletin, 29(4): 1097-1104.

[35] Munir, A.A., 1996. A taxonomisc review of lantana camara L. and L. montevidensis (Spreng) Briq (Verbenacae) in Australia. Journal of the Adelaide Botanical Gardens, 17: 1-27.

[36] Sinha, S., Sharma, A., 1984. Lantana camara L. A review. Feddes Repertorium, 95: 621-633.

[37] Mathur,G., Mohan, R., H.,Y., 1986. Floral biology and pollination of *Lantana camara*. Phyto-morphology, 36: 79-100.

[38] Barrows, E., 1976. Nectar Robbing and Pollination of Lantana camara (Verbenaceae). Biotropica, 8(2): 132-135.

[39] Sahu,A.K., and Panda,S., 1998. Population dynamics of a few dominant plant species around industrial complexes, in WestBengal, India. Journal of the Bombay Natural History Society. 95: 15-18.

[40] Merow, C., Smith, M., and Silander, J., 2013. A practical guide to MaxEnt for modeling species' distributions: What it does, and why inputs and settings matter. Ecography, 36.

[41] Upadhyay, S., Roy, A., Ramprakash, M., Idiculla, J., Kumar, A.S.,Bhattacharya, S., 2017. A network theoretic study of ecological connectivity in Western Himalayas. Ecological Modelling, 359: 246-257.

[42] Newman, M. E. J., 2003. The structure and function of complex networks. SIAM Review, 45(2): 167–256.

[43] Bonacich, P., 1987. Power and centrality: A family of measures. American Journal of Sociology, 92(5): 1170–1182.

[44] Freeman, L. C., 1977. A set of measures of centrality based on betweenness. Sociometry, 40(1): 35–41.

[45] Marchiori, M., Latora, V., 2000. Harmony in the small-world. Physica A: Statistical Mechanics and its Applications, 285(3–4): 539-546.

[46] Watts, D. J., and Strogatz, S. H., 1988. Collective dynamics of 'small-world' networks. Nature, 393: 440–442.

[47] Humphries, M. D., and Gurney, K., 2008. Network small-world-ness: a quantitative method for determining canonical network equivalence. PloS one, 3(4):e0002051.

[48] Newman, M. E. J, 2002. Assortative mixing in networks. Physical Review Letters, 89.

[49] Newman, M. E. J., and Girvan, M., 2003. Finding and evaluating community structure in networks. Physics Review E, 69(2).

[50] Newman, M. E. J., 2003. Fast algorithm for detecting community structure in networks. arXiv:cond-mat/0309508v1.